\documentclass{article}

\usepackage{arxiv}

\usepackage[utf8]{inputenc}
\usepackage[T1]{fontenc}
\usepackage{url}
\usepackage{booktabs}
\usepackage{amsfonts}
\usepackage{nicefrac}
\usepackage{float}
\usepackage{microtype}
\usepackage{graphicx}
\usepackage{caption}
\usepackage[super]{nth}
\usepackage{tabularx}
\usepackage{pdflscape} 
\usepackage{hyperref}
\hypersetup{colorlinks,citecolor=blue,urlcolor=blue}

\usepackage{natbib}
\newdimen\origiwspc
\newdimen\origiwstr

\newenvironment{fontspace}[2]
{\par
    \origiwspc=\fontdimen2\font
    \origiwstr=\fontdimen3\font
    \fontdimen2\font=#1\origiwspc
    \fontdimen3\font=#2\origiwstr
}{\par
  \fontdimen2\font=\origiwspc
  \fontdimen3\font=\origiwstr
}

\title{{\em Pandemic Populism:} \\Facebook pages of alternative news media and the Corona crisis -- a computational content analysis}

\author{
  Svenja~Boberg* \\
  Department of Communication\\
  University of Münster\\
  Münster, 48143 Germany \\
  \texttt{svenja.boberg@uni-muenster.de} \\
\And
 Thorsten Quandt \\
  Department of Communication\\
  University of Münster\\
  Münster, 48143 Germany \\
  \texttt{thorsten.quandt@uni-muenster.de} \\
 \And
 Tim Schatto-Eckrodt \\
  Department of Communication\\
  University of Münster\\
  Münster, 48143 Germany \\
  \texttt{tim.schatto-eckrodt@uni-muenster.de} \\
  \And
  Lena Frischlich \\
  Department of Media and Communication\\
  LMU Munich\\
  Munich, 80538 Germany \\
  \texttt{lena.frischlich@uni-muenster.de} \\
}

\begin{document}
\maketitle

\begin{abstract}
The COVID-19 pandemic has not only had severe political, economic, and societal effects, it has also affected media and communication systems in unprecedented ways. While traditional journalistic media has tried to adapt to the rapidly evolving situation, alternative news media on the Internet have given the events their own ideological spin. Such voices have been criticized for furthering societal confusion and spreading potentially dangerous “fake news” or conspiracy theories via social media and other online channels. The current study analyzes the factual basis of such fears in an initial computational content analysis of alternative news media's output on Facebook during the early Corona crisis, based on a large German data set from January to the second half of March 2020. Using computational content analysis methods, reach, interactions, actors, and topics of the messages were examined, as well as the use of fabricated news and conspiracy theories. The analysis revealed that the alternative news media stay true to message patterns and ideological foundations identified in prior research. While they do not spread obvious lies, they are predominantly sharing overly critical, even anti-systemic messages, opposing the view of the mainstream news media and the political establishment. With this \emph{pandemic populism}, they contribute to a contradictory, menacing, and distrusting worldview, as portrayed in detail in this analysis.

\end{abstract}


\section{Introduction: Corona Crisis, Disinformation, and Pandemic Populism}
When the first case of a new type of respiratory infection were reported to the World Health Organization (WHO) on December 31, 2019 \citep{who_novel_2020}, this disease seemed to be far away and insignificant to most people and the media public around the world. The outbreak was supposedly domestic, limited to Wuhan, China, and many expected it to be controllable and short-lived. Early concerns about a potential global pandemic and severe consequences for affected individuals and societies on a global level were first dismissed as panic-mongering, and the new disease was frequently compared to the seasonal flu both in media reporting and online discussions \citep[for a typical media example, see][and for online discussions, see some of the Corona Reddit threads]{henry_coronavirus_2020}. These assessments were obviously premature, as subsequent developments moved in a different and rather dramatic direction. Just a few weeks later, on March 11, the WHO declared the new disease (then named COVID-19) a global pandemic \citep{ghebreyesus_who_2020}, as the virus (also called SARS-CoV-2) spread at a rapid rate across the planet with large infection numbers and a worrying death toll in many countries. Most of the affected countries reacted with drastic measures to limit the spread of the virus -- with unprecedented restrictions to public life, trade and business, transportation, even small group or individual interactions, and 24/7 curfews with severe penalties for trespassing.

In parallel to the global spread of the virus itself, another sort of pandemic developed—viral and potentially dangerous as well—conspiracy theories, misleading rumors, disinformation, and treacherous lies were shared via online media, including all the larger social media, video, and messenger platforms (i.e., Twitter, Facebook, Instagram, WhatsApp, Youtube, etc.). Scientifically unfounded speculations on potential causes and cures made the rounds, causing confusion and risky behavior among people who followed misleading and false recommendations \citep{phillips_coronavirus_2020}. It was not only anonymous offenders who contributed to public confusion, though. So called “alternative news media” \citep{holt2019}, who position themselves as a counter-force to traditional news outlets, were suspected of spreading such rumors and misleading information, often for political reasons and based on their self-defined “oppositional,” heterodox, and populist perspectives. Some politicians, even in liberal democracies, called for legal action against such malicious disinformation \citep[see][for Germany]{medick_spd-innenminister_2020},  and the WHO reacted with a “myth buster” information site uncovering dangerous rumors and lies about the disease \citep{who_coronavirus_2020}.

Despite the public worries and accusations, the actual contribution of alternative news media and populist voices to public confusion in the Corona crisis remains unclear, also due to the difficulties of analyzing some of the high-impact channels with limited or no \emph{application programming interface} (API) access. There were early and important steps in analyzing, for example, Twitter discussions  \citep{chen2020covid}, but for some countries, other channels were more relevant due to their wider reach; in the case of Germany, only 2\% of the population (14+, German speaking) use Twitter on a daily basis, while 21\% use Facebook daily \citep{beisch_ardzdf-onlinestudie_2019}. Unsurprisingly, many alternative news media publishing in German maintain public Facebook pages with a notable audience, reaching large segments of society either directly or via shared posts.

Therefore, we focused on such outlets in this working paper, documenting an initial computational content analysis of Facebook pages of alternative news media in the (early) Corona crisis. Our data set was limited to the case of Germany and the aforementioned entities, and it covered the initial weeks of the pandemic (up until the decision to introduce a generally binding, national prohibition of contact on March 22, which can be seen as a new phase of the crisis). However, this working paper is part of an ongoing research effort, so subsequent publications are planned to cover the Facebook activities of traditional news media as a contrast to the present paper and to further extend the analysis period. Our aim is to offer substantial data on the public debate of Corona in Germany so that we can move beyond the stage of speculation on who is contributing to such a debate in a potentially risky way. Furthermore, we hope to add to the international research efforts in untangling the communicative confusion that developed parallel to the pandemic and to uncover generalizable patterns of \emph{pandemic populism} that seem to be co-evolving with the spread of the virus itself.

\section{Alternative News Media}\label{alternativemedia}
\subsection{Definition}
Alternative and oppositional voices to professionalized mainstream reporting have always existed, but the global triumph of digital media has substantially eased access to the—now digitized—public spheres. In this environment, which is characterized by diminished authoritative control over information flows, paralleled by a growth of observability in general and paired with decreasing trust in representative institutions \citep{bennett2018}, alternative information channels are flourishing and a vivid ecosystem of an alternative news media prospers.

Alternative news media position themselves as the “corrective of ‘traditional’, ‘legacy’ or ‘mainstream’ news media in a given sociocultural and historical context” \citep{holt2019}.
They stage their “alternativeness” on different levels: by forming alternative media networks \citep[see also the German right-wing populist “free media association,” ][]{ErsteKonferenzFreien2019}, by relying on alternative production routines, such as those described for “critical media” \citep{fuchs2010}, by citing voices usually not heard in the mainstream, by publishing alternative topics, or by employing diverging epistemologies. Often, this kind of alternative content aims at shaping public opinion according to an agenda that is perceived as “being underrepresented, ostracized or otherwise marginalized in mainstream news media” \citep{holt2019}. Following this logic, many alternative news media sites position themselves as an explicit counter-force or watchdog to “mainstream news media,” which also explains a somewhat paradox self-perception: while they need to differ from the mainstream by definition, they are also strongly tied to it, like a mirror image.

\subsection{Alternative News Media and Populism}\label{literaturepopulism}
Although alternative news media are not a new phenomenon, the last years have been marked by a growing number of alternative news media transmitting far-right ideas \citep{holtRightwingAlternativeMedia2019,figenschouChallengingJournalisticAuthority2019} as well as an increasing success of outlets closely tied to political agendas like the Russian-based broadcaster \emph{Russia Today (RT)}, or the Falun Gong’s \emph{Epoch Times}.

In Germany, the most successful alternative news media brands are unified in their affinity to populism \citep{muller2019}. Drawing from the theoretical work of de Vreese et al. (\citeyear{devreesePopulismExpressionPolitical2018}), we focus on populism as a communication style. Populist communication combines the transmission of populist content in the sense of a “thin-centered ideology” \citep{muddePopulistZeitgeist2004} that combines the assumption of a homogeneous and pure people with anti-elitist attitudes and the assumption that elites prevent the true reign of the people \citep[see ][]{schulzMeasuringPopulistAttitudes2017}, with a certain communication style, often characterized by sensationalism and strong emotions \citep{wirzPersuasionEmotionExperimental2018}. This theoretical conceptualization also implies that different degrees of populism exist and that populist communication is likely to vary between different media outlets.

Although not all, and maybe not even most, of the articles published on alternative news media sites with an affinity to populism are directly false, these outlets have been repeatedly accused of serving as disinformation tools willingly discarding the truth to promote their ideological aims \citep{schweigerInformierteBurgerIm2017}, spreading conspiracy theories \citep{yablokovConspiracyTheoriesRussian2015} and relocating the media agenda toward the right \citep{benklerStudyBreitbartledRightwing2017}. Such fact-bending behavior can be particularly problematic in the context of the Corona pandemic where crucial medical information and institutional statements are distributed via mainstream media. The European External Action Service’s East StratCom Task Force (\citeyear{eeasDisinformationCoronavirusShort2020})
reported that they were “witnessing a substantive amount of both misinformation and disinformation spreading on- and offline,” and their website presents multiple examples of disinformation spread by alternative news outlets.

\subsection{Reach of Alternative News Media}
Despite all these concerns, the actual numbers of active consumers of contemporary alternative news media is rather low in most Western democracies. The latest Reuters Digital News Report \citep{Newman2019} found that although 44\% of US citizens knew the alternative flag-ship Breitbart, only 7\% had used it in the previous week. Numbers for other outlets and in countries like the UK, France or, of relevance for the current study, Germany are even lower \citep{hoeligReutersInstituteDigital2019}. 

Nevertheless, single alternative news outlets do create a substantial amount of online activity in Germany. Two out of ten right-wing populist alternative news media studied by \cite{heftBreitbartComparingRight2019} were found to be among the top 300 German websites based on their Alexa.com scores. It is thus of little surprise that the amount of occasional or incidental consumption is rather high. In a quota-survey, 8.2\%–15.3\% of Germans above the age of 18 reported at least occasional exposure to alternative news media with an affinity to populism \citep{muller2019}. Moreover, despite their often-untrustworthy reporting, the media-trust panel of the German University of Mainz found that 42\% of Germans perceived news on alternative news websites to be rather or very trustworthy \citep{ziegeleLugenpresseHysterieEbbt2018}. 

\section{Research Interest and Research Questions}
Following the above-mentioned argument, we focus on alternative news media and their Corona-related online (and in particular: Facebook) activities. While there is a public suspicion and fear that they will contribute to confusion and disinformation, and as a result, negatively impact societal functioning, there are deviating findings that put such statements into perspective, claiming relatively low relevance. What’s more, not much is known about what kind of content they distribute in the crisis; before claiming effects of messages, one should know the messages themselves to develop ideas on their potential to even elicit effects. Therefore, we will apply computational methods of content analysis to qualify the message structure of alternative news media in the Corona crisis, asking a few deliberately simple, but essential questions:

\hspace{15pt}[RQ1a] How broad is the reach of the alternative news media’s Corona coverage on Facebook?

\hspace{15pt}[RQ1b] How many interactions do they evoke?

\hspace{15pt}[RQ2] What are the central topics of their coverage of the Corona crisis?

\hspace{15pt}[RQ3] Who are the most prominent actors in their coverage?

\hspace{15pt}\hangindent=15pt[RQ4] Do the messages of alternative news media include fake news or conspiracy theories, as identified by fact checking entities?

Naturally, these questions are not understood in isolation of other research efforts on the Corona crisis. In a subsequent working paper, we will focus on traditional news media and their Facebook activities, asking the same questions to allow for comparisons. However, for the time being, the above questions will give initial insight into how alternative news media cover the crisis and whether we can find the assumed populist bias in terms of sources, topics, and actors.

\section{Method}
\label{sec:method}

\subsection{Sample and Data}
In the current study, we focus on German alternative news media sites on Facebook over the course of the early Corona crisis, from the beginnings up to the phase of pandemic spread and significant impact in Germany, in particular. The data set starts on January 7, 2020, when Chinese authorities first identified a new type of Corona virus to be the culprit behind the disease  \citep{who_novel_2020}, and ends on March 22, 2020. At this point, Europe registered more infected individuals than any other region, and Germany had already introduced measures to restrict public life, closed borders and, in some places, imposed curfews or strict social contact regulations. Access to the Facebook pages and the crawling of their public content was enabled via \emph{CrowdTangle}, which in turn was made possible through participation of the research team in Facebook’s transparency initiative. CrowdTangle tracks the public (non-targeted, non-gated) content from influential or verified Facebook profiles and pages. It does not provide insights into private user content.\footnote{For details see \url{https://help.crowdtangle.com/en/articles/2541882-faq-general-crowdtangle-questions} }

The basis of the crawling is the previous identification of 71 alternative news media sites in Germany on the basis of the literature \citep [particularly][]{bachlAlternativeMediaSources2018,heftBreitbartComparingRight2019, puschmannInformationLaunderingCounterpublics2016, schweigerInformierteBurgerIm2017}, and our own research in the respective media ecosystem \citep[e.g.,][]{frischlichMainstreamAlternativeCoorientation}, as well as the identified media's public self-descriptions, own link analyses, and audience overlap scores (via Alexa.com). Some of these are operating from Austria or Switzerland but cover issues of relevance for Germany as well and are therefore included in the analysis (which is, however, explicitly not a full analysis of the situation in Austria or Switzerland). Forty of these alternative news media had a Facebook presence at the time of the data collection; eight were inactive during the analysis period. We crawled the remaining 32 to examine these media’s reporting on the new corona virus.

In parallel to the analysis of the alternative news media sites, we crawled the Facebook presence of the leading “mainstream” newspapers for a subsequent analysis using an updated version of a previously collected overview of German mainstream newspapers  \citep{frischlichCommentSectionsTargets2019}. The procedures for conceptualizing this list differed from the approach used to identify the alternative news media. Following a multi-level triangulation approach, we first identified all online media with more than 500,000 unique users using the AGOF database \citep{agofMonatsberichteJanuar20192019}; then, we trimmed the list and excluded all non-journalistic outlets \citep[for a similar approach, see][]{steindlJournalismusDeutschlandAktuelle2017,weischenbergJournalismusDeutschland2005}, and identified online newspapers within this list based on the literature \citep [particularly, ][]{buhl_observing_2018, schutzDeutscheTagespresse20122012} and our own investigations. Through this analysis, we identified 80 national and regional online newspapers. Seventy-eight of them hosted at least one active Facebook page at the time of data collection. Only the main page for each newspaper was incorporated in our data base. In order to check for circulating fake news stories or conspiracy theories in the examined time period, we also collected all public posts of the most prominent German fact-checking sites, \emph{Mimikama} and \emph{Correctiv}.

The crawled CrowdTangle data set includes all the posts during the respective time period with a total of 116,994 posts, broken down into 100,432 mainstream media posts, 15,207 alternative news media posts, and  1,355 fact-checking posts. The full list of pages is available as an ancillary file on \href{https://arxiv.org/src/2004.02566v2/anc/supplements_sample_full_media_list.csv}{arXiv.org}. CrowdTangle offers the typical set of meta data like timestamps, page likes, engagement metrics (number of likes, shares, and comments), and information on linked URLs in the posts (including embedded headlines and teaser texts of the linked sites or articles). For the analysis, we filtered the Corona-related posts based on regular expressions covering different spellings and technical terms for “Corona” and “COVID-19” and topic-related terms like “epidemic”, “pandemic,” and “quarantine” or often-used hastags like “\#flattenthecurve”, “\#washyourhands” and “\#stayhome.” As a result, our Corona-related sample consisted of 2,446 alternative news media, 18,051 mainstream, and 282 fact-checking posts.

\subsection{Analysis}
\subsubsection {Reach}
To determine the reach of the single media outlets, we used the number of page likes provided by CrowdTangle. The number of followers was not retrieved historically but corresponds to the number of followers at the date of the data crawling. The number of likes, shares, comments, and reactions was also provided by CrowdTangle. These metrics are added (unweighted) and summarized in the variable “total interactions.”

\subsubsection {Preprocessing}
In order to prepare the data for analysis, a series of common preprocessing steps was applied \citep{gunther_word_2016}, including removing fragments of html-markup, URLs, mentions and hashtags, punctuation, and stopwords. To handle the ambiguous use of names, the named entities mentioned in the posts were annotated using the German classifiers of the Stanford CoreNLP Toolkit \citep{manning-EtAl:2014:P14-5} and standardized manually. This involved names of persons (e.g., “Mrs. Merkel” and “Angela Merkel”), organizations (e.g., “the green party” and “Alliance 90/The Greens”), or states (e.g., “United States” and “USA”). We also made an effort to assign political organizations to the respective context (e.g., “government”, “German government”, or “Austrian government”).

\subsubsection {Analytical Approach}
The exploration of the message composition with previously unknown topics can be done using inductive methods that categorize the material according to its underlying thematic structure, for example, on the basis of a statistical analysis of the probability, coherence, and similarity of word patterns. This class of methods, like document clustering or topic modeling, does not require predefined dictionaries, but still requires some interpretation and decisions of the researcher to find a fitting structure that is not only mathematically acceptable, but also sound when it comes to the meaning of the generated categories \citep{gunther_word_2016}. To identify the underlying thematic focus of alternative media, we used \emph{latent Dirichlet allocation} \citep[LDA; ][]{blei2003latent}, an unsupervised learning algorithm that can discover latent topics inductively based on patterns of terms occurring together. The algorithm can detect to what extent each word of the corpus characterizes the respective topic ($\beta$) and how far each topic is present in each document ($\gamma$), so each post can be a mixture of more than one topic \citep{gunther2017communication}. Before estimating the topic model, the number of topics (\emph{k}) has to be predefined. To find the ideal \emph{k}, a series of 2 to 50 topic models were computed based on training and test samples using the LDA function of the R topic models package \citep{hornik2011topicmodels}. We found that the model of \emph{k} = 11 had a large increase in predictive power. This finding and the corresponding parameters have also been confirmed using the R ldatuning package  \citep{nikita2016package}. The eleven topics were characterized by inspecting the top terms and most representative documents for each topic. The interpretation was validated by word and topic intrusion tests \citep{NIPS2009_3700}, checking whether an independent coder could identify terms and topics that were randomly assigned, but not representative for the respective topics. 

To portray the context of the most prominent actors as proposed in RQ3, we conducted a co-occurrence analysis  \citep{bordag2008comparison}, by counting the frequency of word a and word b appearing in the same post. The co-occurrence count was computed on the whole sample and then filtered by the 20 most frequent named entities (as annotated during preprocessing, see Table \ref{tab:top_20_ner_actors}). We focused on persons (e.g., Angela Merkel), institutions (e.g., the German government) and organizations (e.g., the CDU party) in the analysis and excluded locations like states and cities deliberately (as they are not core to our RQs). The co-occurrence network was then visualized as a force-directed graph using the R-package igraph \citep{csardi2006igraph}. 

In order to identify conspiracy theories and fake news stories (RQ4), the Corona-related posts of fact-checking (\emph{N}~=~282) sites were reviewed. It is noteworthy that fact checkers not only monitor alternative news channels but also address Internet scams, online chain letters, incidents of offline fraud (e.g., tricksters posing as members of the health department), and misleading headlines or pictures, which can be described as poor reporting of mainstream journalism. However, we focused only on debunked “fake news” and conspiracy theories in this paper. Often there were related versions of the same story, only differing in singular aspects like the suspected culprit of the conspiracy. While the fact checkers have taken up many of these different forms (hence the number of 282 Corona-related posts), we have summarized them in the overarching narratives, resulting in four conspiracy and fake news stories (for a detailed description see Section \ref{fake_con}). We constructed a dictionary for each story and applied it to the sample in order to check whether the respective stories were covered in the alternative news media. This is admittedly a rather direct and simple, but nevertheless valid, procedure since the stories were always connected either with concrete persons (e.g.,“Wolfgang Wodarg” or “Bill Gates”) or specific keywords (e.g., “laboratory virus”).

It needs to be noted that we are very aware of the problems tied to the concept of fake news as discussed elsewhere \citep{quandtFakeNews2019}, but we use it deliberately here, as it is defined as a category of fact checkers denoting false and fabricated news items (and not, for example, in the sense of a political allegation, as frequently used by US President Donald Trump). In the context of our study, it also has an advantage over the widely-used concept of disinformation, as it does not imply intentionality, which we cannot infer from a content analysis alone. That said, we would like to point out that the use of the term is a very narrowly defined one in the context of this article.

\section{Results}

\subsection{Reach and Total Interactions}\label{reach}

\begin{table}[h]
 \caption{Media types}
  \centering
  \begin{tabularx}{\textwidth}{lr rr rr rr rr}
    \toprule
     & & \multicolumn{2}{c}{Page likes} &
     \multicolumn{2}{c}{Posts} &
     \multicolumn{2}{c}{Interactions} & \multicolumn{2}{c}{Shares} \\
     \cmidrule(lr){3-4} \cmidrule(lr){5-6} \cmidrule(lr){7-8} \cmidrule(lr){9-10}
    Media Type & Pages & Total & Per page & Total & Per page & Total & Per post & Total & Per post \\
    \midrule
    Alternative  & 32 & 1,653,208 & 51,663 & 2,446 & 76 & 589,534 & 241 & 197,401 & 81 \\
    Mainstream   & 78 & 16,779,317 & 215,119 & 18,051 & 231 & 5,732,155 & 318 & 1,472,648 & 82 \\
    \bottomrule
 \end{tabularx}
 \label{tab:media_systems}
\end{table}

\begin{figure}[h!]
  \centering
  \includegraphics[width=1.0\textwidth]{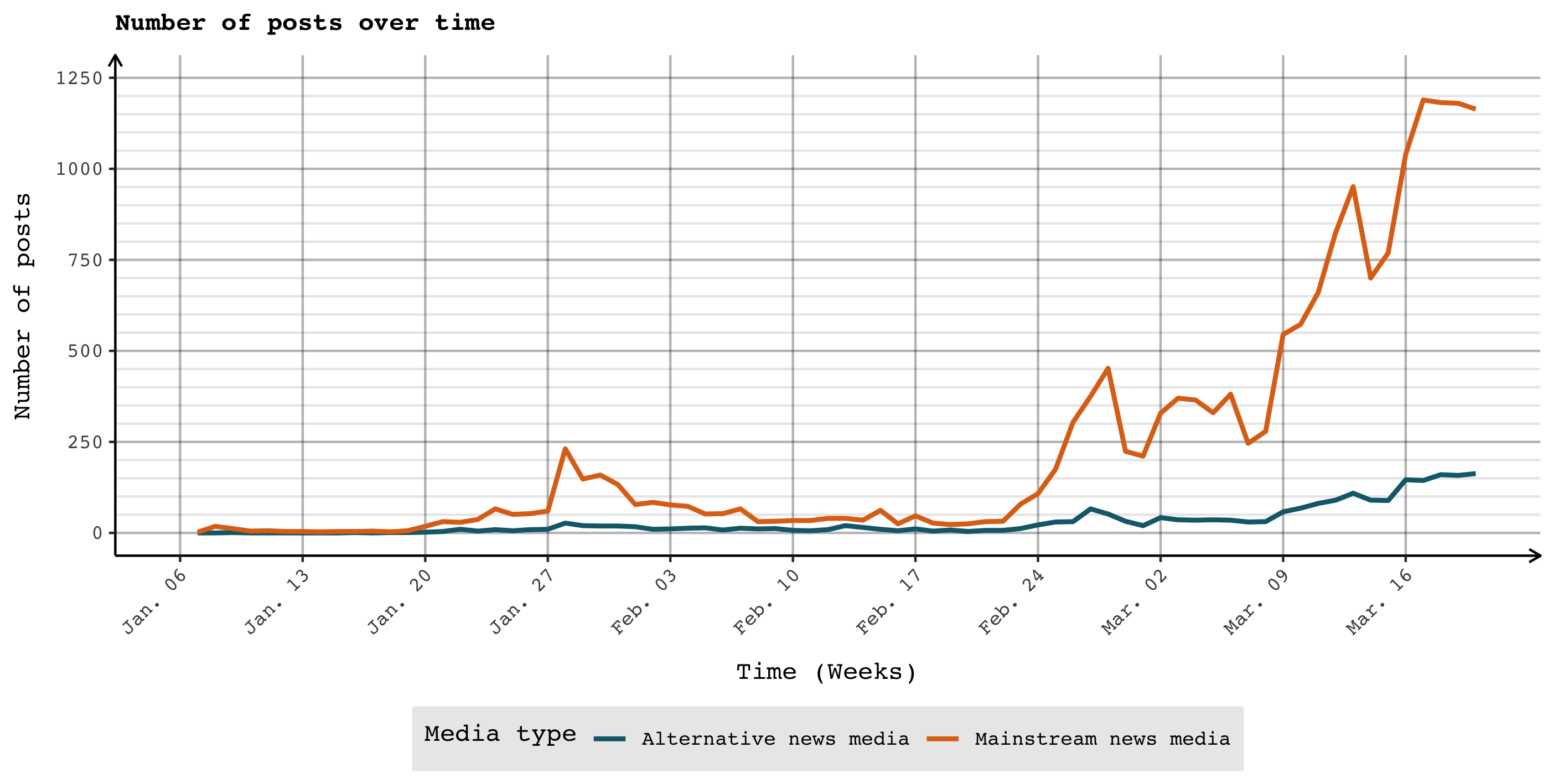}
  \caption{Number of posts over time}
  \label{fig:number of posts over time}
\end{figure}

In a first step, we compared alternative and mainstream media by their output, potential reach, interactions, and shares. Unsurprisingly, we found that the traditional mainstream media had a much higher output of posts referring to the Corona crisis, both in total and by the average number of posts per page (see Table \ref{tab:media_systems}). The first reported infection in Germany on January 28 \citep{tagesschaude_deutscher_2020} and the complete lock-down in Italy’s Lombardy region at the end of February \citep{cnneditorialresearchCoronavirusOutbreakTimeline2020} seem to have influenced the amount of coverage in both media types, but with a more noticeable effect on the mainstream news media due to their higher absolute number (see Figure \ref{fig:number of posts over time}). The time-based graph shows two spikes around these dates, a typical pattern of event-based bursts \citep{buhl_bad_2019, buhl_observing_2018}. During March, the coverage in the mainstream media grew massively, seemingly mimicking the exponential spread of the virus in society—a reaction to the now strongly perceivable medical, political/regulatory, and societal effects reflected in the increased reporting. While the overall number of posts of the alternative news media was much lower, they also followed the same growth pattern (see Section \ref{emergence}).

The number of likes a Facebook page has is an indicator of the number of users who receive updates about the page in their personal feed, that is, a page's potential reach. Comparing the media types in terms of their potential reach, the mainstream news media in our sample had more than ten times the number of likes compared to the alternative outlets. This is partially due to the higher number of mainstream outlets in the sample. On a per-page basis, mainstream media outlets had only slightly more than four times the number of likes.

\clearpage

\begin{figure}[h!]
  \centering
  \includegraphics[width=1.0\textwidth]{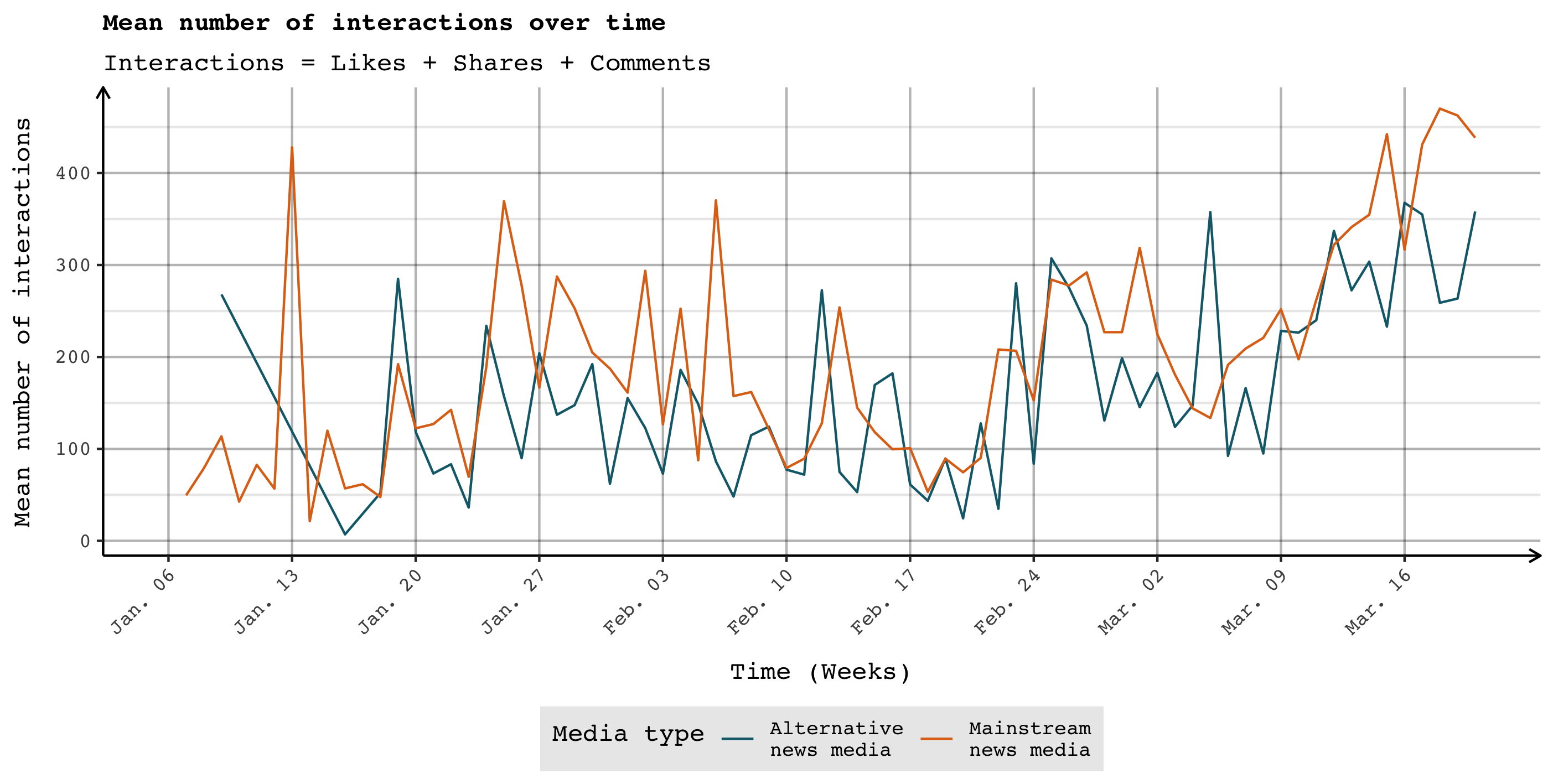}
  \caption{Mean number of interactions over time}
  \label{fig:mean number of interactions over time}
\end{figure}

In terms of \emph{total} interactions (i.e. the number of likes, comments, emojis and shares a post received), the mainstream news media generated a larger engagement than the alternative news media (see Table \ref{tab:media_systems}). However, comparing the \emph{average} interactions per post revealed that, on average, the alternative outlets had only 25\% fewer interactions than their mainstream counterparts (\textit{t}(20495)~=~-3.93, \textit{p}~<~.001). This rate remains relatively constant over time, with a growth of overall interactions beginning at the end of February (see Figure \ref{fig:mean number of interactions over time}), which also parallels the growth of coverage (and plausibly societal interest).

Mainstream news media also produced a larger share of the best performing posts in terms of total interactions. Among the top 1\% of posts, only 4.9\% originated from one of the alternative outlets. All of them were posted by either \emph{RT Deutsch}, \emph{Junge Freiheit}, or the \emph{Compact-Magazin}. The alternative news media performed slightly better when only the most frequently shared posts were inspected. In the top 1\% of the most frequently shared posts, 11.27\% originated from one of the alternative news media outlets.

In summary, the mainstream news media had a higher output, generated more interactions both in total and per post, and had a higher potential reach. This is not surprising, given the differences in production resources and staff: Alternative news media are often small and not as well funded as their mainstream counter-parts. Also noteworthy is the average number of interactions per post not differing as greatly as the difference in potential reach would suggest. Our analysis showed that even though the mainstream media outlets were dominating the overall information environment on Facebook, single posts by alternative news outlets reached a comparable number of interactions and shares to those published by mainstream outlets, indicating that these might have developed considerable impact in their respective audience group.

\subsection{Topics}\label{topics}
\subsubsection{Identification of main topics}
The previous analysis answered the more formal RQ1a and RQ1b regarding reach of and interactions with alternative news media vis-à-vis mainstream news media, but did not focus the actual content of the messages beyond pointing at the share of Corona-related content. With the help of topic modeling as a more nuanced method, it is possible to take a deeper look at the composition of the material and see what the alternative news media were covering during the analysis phase. An eleven-topic model emerged as the ideal solution to describe the underlying topical structure (see Table \ref{tab:TM_description}). 

\clearpage

\begin{table}[p]
 \caption{Topic description}
  \centering
  \begin{tabular}{
  >{\raggedright\arraybackslash}p{0.6cm}%
  >{\raggedright\arraybackslash}p{2cm}%
  >{\raggedright\arraybackslash}p{2cm}%
  >{\raggedright\arraybackslash}p{7cm}%
  >{\raggedright\arraybackslash}p{1cm}%
  >{\raggedright\arraybackslash}p{1cm}%
  }
    \toprule
    Topic & Description & Top terms & Example & Mean $\gamma$ & N, where $\gamma$ > 0.5    \\
    \midrule
    1   & Number of infected and countermeasures    & Germany, Italy, China, number, cases  & “The number of coronavirus infections is skyrocketing. Until Wednesday evening 102 more than in the last official report.” (MMnews)  & 0.18 & 403 \\
    2   & Failure of governments in crisis management   & Wuhan, China, SARS, outbreak, pandemic & “Germany is well prepared!? The number of political chatterboxes is huge. Nothing can go wrong -- can it?” (ScienceFiles) & 0.11 & 242\\
    3   & Economic crisis due to misguided political action   & crisis, economy, German, pandemic, consequences & “Leading German economists have rejected so-called helicopter money in the fight against the economic consequences of the corona crisis. At the same time, Finance Minister Olaf Scholz (SPD) has promised further aid. Economists, however, predict severe recession.” (Junge Freiheit)  & 0.11 & 240\\
    4   & Measures are not reasonable    & Germany, crisis, pandemic, measures, politics & “Anyone who observes the reaction of Austrian politics to the corona epidemic has to realize every sentence is textbook PR. […] Several mistakes can be pinpointed - but they are ignored by the media.” (Andreas Unterberger)  & 0.07 & 138\\
    5   & Chaotic crisis management & China, Russia, crisis, panic, Europe, pandemic &  “Moscow: Supermarkets still well stocked, situation relatively relaxed. Panic has taken hold in many countries with the Corona virus. Goods such as soap, disinfectants and toilet paper are completely sold out.” (RT Deutsch)  & 0.10 & 198 \\
    6   & Fear and panic    & Please, fear, contribution, share, Germany & “DISCOVERED: Corona displaces climate change +++ The same pattern of argumentation as for climate change is now being used in numerous media -- we recall a special trick... Please SHARE the article” (Politaia) & 0.07 & 139\\
    7   & Worst case scenarios    & Trump, foto, USA, canceled, Germany & “Fear of \#Coronisation -- \#UK is experiencing an U-turn in the government’s fight against the \#Coronavirus. Business organisations fear state nationalization of companies.” (Junge Welt) & 0.08 & 167 \\
    8  & National cohesion vs. failure of individuals & Merkel, crisis, EU, CDU, Germany & “It is well known that left-wing extremists use every opportunity to overthrow this state. Now in the Corona crisis a group called ‘Revolutionary Antibodies’ is calling for plundering.” (Junge Freiheit) & 0.11 & 237 \\
    9  & Corona and migration & migrants, EU, quarantine, Turkey, Erdoğan & “Migrants have violently tried to break the quarantine of their facility. […] This aggressive group of migrants poses an imminent threat.” (Achse des Guten) & 0.06 & 109 \\
    10  & Clickbait & Iran, quaratine, measures, positive, infected & “To prevent the penguins from getting bored during the quarantine, the management of the aquarium in Chicago decided to organize an excursion for them.” (Sputnik) & 0.07 & 146 \\
    11  & Corona as a buzzword & Hints, day, week, overview, AfD & “Against socialism -- Our reading list for the quarantine” (ScienceFiles) & 0.04 & 82 \\
    \bottomrule
    \multicolumn{6}{l}{\emph{Notes:} LDA with \emph{k} = 11, \emph{$\alpha$} = 0.05, sampling method = Gibbs} \\
  \end{tabular}\label{tab:TM_description}
\end{table}

\clearpage

A first inspection of the resulting topics revealed some that more or less fit traditional mainstream media reporting, such as Topic 1, which includes the number of infections and the measures to deal with the pandemic. Others topics clearly fit alternative news media’s heterodox, anti-establishment perspective or right-wing position, like Topic 9, which refers to an anti-migration narrative linked to Corona. Still others used the pandemic as clickbait or a buzzword to simply draw attention, like Topic 11, which also included links to products and services.

A closer look uncovered patterns of coverage in alternative news media offering some clues about how the information is typically framed (via the top terms in the respective topics). The topic with the highest count of messages containing such content to a large extent (\emph{n}~=~403) was the primarily factual coverage of infection numbers, deaths, quarantine, and curfew measures, both in Germany and elsewhere (Topic 1). It does not differ much from what one would expect from mainstream media and also refers to official sources like regional health departments.

Topic 2 deviates from the expectations for traditional news reporting by offering often opinionated, system-critical messages. It unites the failures of national and international authorities’ actions to deal with the growing pandemic, partially in a sarcastic tone (for per-topic message examples, see Table \ref{tab:TM_description}). This topic also includes critical comments on the German focus on the scandalous election in the state of Thuringia or activism against climate change, instead of preparing for the "much more relevant" pandemic. It is necessary to note that this is somewhat contradictory to some of the alternative media’s own preoccupation with promoting the nationalist Alternative für Deutschland (AfD) party, which played a major role in the scandal around the election in Thuringia, and characterizing activism against climate change being the misleading idea of children from affluent, left-wing families or some sinister conspiracy to destroy Western economies.

The third topic focused on the economic consequences of the pandemic and criticized the actions of leading politicians, with some alternative news media predicting a major crash of the markets and asking whether the economic damage was even worse than the damage to health caused by the virus. This line of argumentation is also reflected in Topic 4, extending the notion that political measures are excessive and not helping the cause. It portrays the actions as symbolic politics and driven by panic. This criticism is supported with reference to the statement that Corona is no worse than a normal flu, and with indignation at SPD politician Boris Pistorius’ attempt to sanction Corona-related disinformation.

The fifth topic is somewhat like the second one as it covers the situation in various countries, pointing out that some nations are doing quite well, while others sink into chaos. For example, the Russian-friendly channel, RT Deutsch, praises the relaxed situation in Moscow with well-filled supermarkets and contrasts this with the panic elsewhere to insinuate systemic advantages in the crisis. The sixth topic, again picks up the panic-mongering element, but condenses it to a criticism of a fear-inducing system of (mainstream) media and politics per se. For example, it speculates on recurring patterns and similarities between the climate change debate and the pandemic, implying some strategy behind it. This has the components of a conspiracy theory and again refers to the already existing narration of climate change activism being either misguided or part of some societal steering. Interesting enough, Topic 7 serves exactly the panic mongering and fear appeals criticized in the sixth one, as it sketches worst-case scenarios and state of emergency fantasies (with links to the situation in the United States and frequent mentions of President Donald Trump), including military action, nationalization, and other drastic measures.

The eighth topic focused on societal cohesion and the reckless actions of individuals who harm the community. For example, left groups are pilloried for using the crisis as a tool to undermine the state structure. Also, in this topic, political representatives, like Chancellor Angela Merkel or the Bavarian state president Markus Söder, appear prominently. In contrast to the other topics, they are not entirely portrayed as flawed here. We also found a nationalist, submissive to authority tone here, which also supports the findings of previous research (see Section \ref{alternativemedia}). This is somewhat paradox, as some messages and topics (like this one) imply loyalty to the nation and its representatives, while others completely state the opposite.

Topic 9, again, mixed a long-term narrative of the right-wing alternative news media in Germany with the current pandemic; it links the Corona crisis with a typical anti-migration and anti-refugee stance, which often includes references to Turkey and its refugee camps and the situation in Greece. Turkish President Erdoğan is mentioned frequently in this topic as he announced letting refugees from Turkish camps travel to Europe during that time period. A decision which was framed as a threat to further the spreading of the virus by some alternative news media (see Section \ref{actors}). In that sense, this topic corresponds to the alternative media’s pre-crisis worldview \citep{heftBreitbartComparingRight2019} and can be seen as an extension of the respective narrative. The crisis is more or less simply assimilated, and Corona-related news is integrated into that perspective, for example, when contagious migrants are portrayed as a threat to German citizens.

The last two topics are different from the previous ones because they are only partially focused on the pandemic; Topic 10 primarily consisted of soft news (like news about animals in the crisis) or highly emotionalized posts. Grosso modo, this is clickbait material. Topic 11 is only present in a relatively small number of posts, and here, Corona only showed up as a buzzword used to raise interest for articles, services, or products. In contrast to the clickbait material, which has at least had some topical link to the crisis, it uses typical Corona-crisis language to grab the attention of the reader and to raise awareness for something virtually unrelated.

Summing up, the topics present in the Corona coverage of alternative news media follow their worldview and familiar narratives, primarily offering a populist, anti-systemic, anti-establishment perspective. This is consistent with the expectation that even in such an extraordinary event the outlets studied here stay true to topical patterns pre-dating the crisis (which is clearly visible in the case of topics that were simply adapted to the crisis, like the critical stance toward migration and refugees or climate change activism).

\subsubsection{Emergence and development of topics over time}\label{emergence}
\begin{figure}[h]
  \centering
  \includegraphics[width=1.0\textwidth]{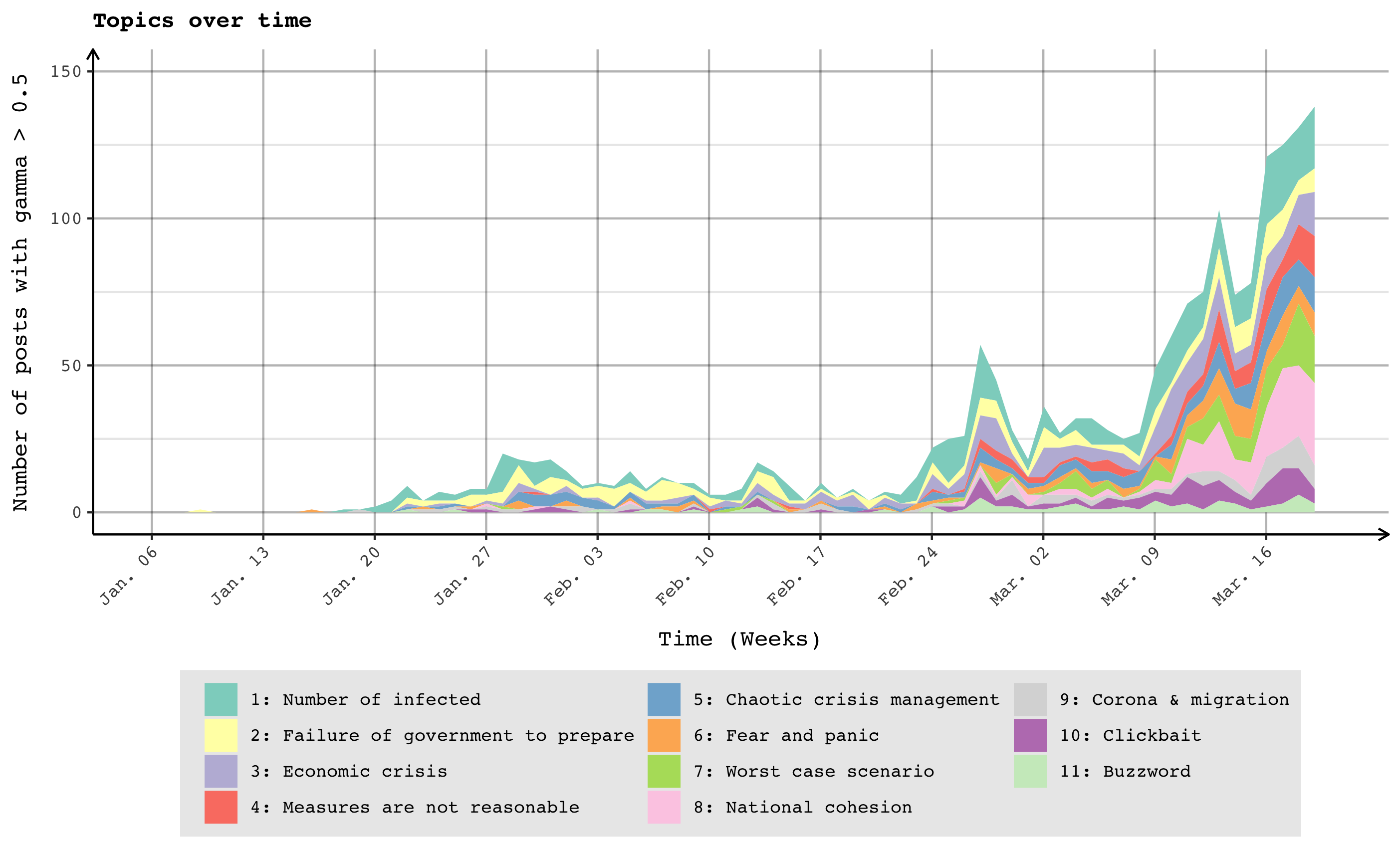}
  \caption{Number of topical posts over time}
  \label{fig:topic posts over time}
\end{figure}

Up to now, we described the topical composition of the overall data set in an aggregated way, irrespective of the publishing dates of the messages. However, as discussed in the analysis of the overall output of posts, the debate developed in multiple stages, and there were several notable spikes or “event bursts” (see Section \ref{reach}). In an additional research step, we, therefore, investigate the emergence and flow of topics by adding the time dimension to the analysis of the previously identified topics. The resulting chart (see Figure \ref{fig:topic posts over time}) reveals the development of alternative media’s overall output to follow the curve of the mainstream media with great accuracy (see Figure \ref{fig:number of posts over time}).

Up to January 20, the virus was essentially a non-issue in the analyzed Facebook debates. With the first Coronavirus deaths in China (officially announced on January 11), reports of infections occurring outside of China (Thailand on January 13 and Japan on January 16), and with the growing number of cases throughout the region, the topic slowly raised interest in alternative news media \citep[for an international event timeline see, for example, ][]{cnneditorialresearchCoronavirusOutbreakTimeline2020}. The first handful of reports focused on infection numbers (Topic 1) and criticism of governments’ lack of preparedness (Topic 2).
 
The overall news output on the topic gained some momentum from January 28 onward, which is the date of the first known infection in Germany \citep{tagesschaude_deutscher_2020}. This case and the subsequent infections were limited to staff of automotive supplier Webasto in Bavaria and the infection chain was traceable, so the situation seemed to be contained \citep{robert_koch-institut_bulletin_2020}. Accordingly, alternative news media paid some attention to the topic but after the initial push, activity remained limited, with some criticism of the lack of preparedness being the dominant reaction.

This situation remained stable until the final days of February. On February 24--25, two persons in the city of Heinsberg and one person in the city of Göppingen were identified as being infected \citep{sohr_erste_2020}. The former is relevant as those two had visited a carnival before being diagnosed and had infected numerous people there, leading to a heavy outbreak of the virus in the area of Heinsberg. The person from Göppingen was found to have been infected during a recent trip to Italy, where the highest infection rates were recorded in the weeks that followed. Starting with these events, the spread of the virus in Germany changed from a contained situation to a more complex one with the unfolding signs of an epidemic. In parallel, the reports on alternative news media grew considerably.

Additionally, the dramatic situation in Lombardy (in the north of Italy) along with Germany’s decision on February 26 and 27 to declare some regions in Italy as risk areas \citep{robert_koch-institut_covid-26_2_2020, robert_koch-institut_covid-27_2_2020} lead to a notable news burst \citep{buhl_bad_2019, buhl_observing_2018}, i.e. a visible spike in curve (as also noted in Section \ref{reach}). These events also led to a change in the composition of topics: The constant criticism of the measures taken by the government (Topic 4) basically started at that time (disregarding some earlier singular posts in that topic category), while the negative evaluation of the crisis management (Topic 5) and the issue of fear and panic (Topic 6) were revived (after having produced some singular posts following the Webasto cases at the end of January and early February).
Interestingly, these events also pushed clickbait material (Topic 10) and the use of Corona-related terms as buzzwords or links to something completely unrelated (Topic 11). The exploitation of the unfolding pandemic for advertising, services, or generating clicks may be ethically questionable, but it follows a simple  “attention economy” \citep{goldhaber_attention_1997, davenport_attention_2001} logic: Anything virus-related is supposed to grab the attention of the users, which can be, in turn, converted into economic value.

After the spike, we see a dramatic drop in posts, the reason possibly as mundane as the fact that February 29 and March 1 was a weekend. Similar drops in output on the weekends can also be seen in subsequent weeks (March 7--8 and 14--15). It is very plausible that many of the alternative news media do not produce content on weekends or have a much-reduced output. While such weekly seasonality is well known in time studies analyzing human work, it is still worth noting that even under such extraordinary circumstances, the effects are quite significant, and some routines seem to remain unchanged.

The final and rather dramatic growth phase starts on March 9, which parallels the equally dramatic events that unfolded from that date up to the end of our analysis phase. That particular Monday was the date of the first two Corona-related deaths in Germany \citep{n-tvde_zwei_2020}. In the days that followed, a growing number of German ski-tourists returning from Austria showed signs of infection, and the Austrian state of Tyrol was declared a risk area by the German Robert Koch Institute on March 13 \citep{robert_koch-institut_lagebericht_13_3_2020}. The Austrian government quarantined some of the popular ski areas on the same day, leading to chaotic scenes of tourists desperately trying to leave the area \citep{willim_quarantanezone_2020, dahlkamp_corona_2020}. In subsequent days, more and more areas and countries were added to the list of risk zones, while the number of infected as well as the death toll were growing at a worrying rate. The troubling developments all over Europe lead to several countries closing their borders and imposing drastic measures to slow the spread of the virus.
Unsurprisingly, we see notable growth in most of the topics during that phase, in parallel to the overall increase in news output. The mushrooming of various doom-and-gloom scenarios (Topic 7) during that dramatic period is especially notable, even more so the emergence and rapid growth of the topic of national cohesion (Topic 8). This can be plausibly explained as a reaction to the perceived threat and, partially, as a “rally around the flag” effect. In the case of right-wing populist media, this does not necessarily extend to the political actors (which are seen as representatives of "the system") but to the nation as the unifying entity. That said, the topic category at least contains a notable number of neutral posts on the actions of leading politicians, which deviates from the norm of overly critical remarks in alternative news media. In particular, the television address of Chancellor Merkel on March 18 \citep{TVanspracheMerkel} and her call to solidarity were welcomed with unusual restraint.

In summary, the time-based analysis reveals some striking patterns of topic emergence, change, and growth. The pandemic spread of the virus is paralleled by an explosive growth of media output, and some key events lead to notable “bursts” and the introduction of new topics to the overall mix of news. Interesting enough, the weekend drops in the production are still quite significant, even at a time of unprecedented crisis coverage—an observation that also holds true for the mainstream news media (see Figure \ref{fig:number of posts over time}). It needs to be noted, though, that these time-based and per-topic analyses have to be interpreted with caution. Despite the overall number of posts in the data set being of a solid size, it is spread out over multiple weeks and across 11 topics, so some of the results—especially in the early phase—refer to a limited number of messages.

\subsection{Actors}\label{actors}
\subsubsection {Top-ranked actors}

As we know from previous research, media often uses personalization, especially in political reporting, to reduce complexity and to tell stories alongside the actions of individuals \citep{Boumans2013, Bennett2012}. Group entities \citep[and even nations, s.][]{cohen_foreign_2012} are similarly portrayed as unified persons, implying their actions as deliberate and coordinated, that is, organization A does something and then organization B reacts. Such reporting scales down complex processes to the level of situational interaction and face-to-face communication, and thus transfers it to the real-world experiences of the audience, which also makes complex processes easier to grasp. It can be argued that this form of reduction is even more relevant in the Corona crisis where the threat comes from a virus that cannot be directly perceived and is not embodied, and as such, remains on a conceptually abstract level (while still being very real in its effect). The analysis of actors, in a broad sense, therefore allows for a deeper understanding not only of who is frequently referred to as an important person or enemy, but also of the construction of the story, and thus, the logics and functioning of the reporting. For the current analysis, we identified the most frequent named entities in the data set (excluding nations, as they were not central to our research interest; see Section \ref{sec:method}). 

\begin{table}[H]
 \caption{Top 20 most frequent named entities}
 \centering
  \begin{tabular}{llr}
    \toprule
    Entity & Notes & \emph{n} \\
    \midrule
    Angela Merkel & Chancellor of Germany & 308 \\
    Bundesregierung & Federal Government of Germany &  138 \\
    Jens Spahn & Federal Minister of Health & 131 \\
    Donald J. Trump & \nth{45} US president & 117 \\
    CDU & Christian-democratic, liberal-conservative political party in Germany & 110 \\
    AfD & Far-right political party in Germany & 58 \\
    Markus Söder & Minister President of Bavaria & 52 \\
    SPD & Social-democratic political party & 49 \\
    Robert Koch Institute (RKI) & German federal government agency and research institute & 48 \\
    Recep Tayyip Erdoğan & President of Turkey & 42 \\
    Emmanuel Macron & President of France & 40 \\
    Friedrich Merz & Former MEP and leader of the CDU/CSU group in the German parliament & 31 \\
    Sebastian Kurz & Chancellor of Austria & 27 \\
    Greta Thunberg & Swedish climate activist & 26 \\
    Peter Altmaier & German Federal Minister for Economic Affairs and Energy & 25 \\
    Die Grünen & Green political party in Germany & 23 \\
    Christian Drosten & German virologist & 23 \\
    Wolfgang Wodarg & Former member of the German parliament and medical doctor & 23 \\
    FDP & Liberal political party in Germany & 21 \\
    Olaf Scholz & German Federal Minister of Finance and Vice Chancellor & 20 \\
    \bottomrule
    \multicolumn{2}{l}{\emph{Notes:} \emph{N} = 421 } \\
  \end{tabular}
  \label{tab:top_20_ner_actors}
\end{table}

Unsurprisingly, the most prominent entities mentioned were the political actors coining the crisis, both national and international (see Table \ref{tab:top_20_ner_actors}). German Chancellor Angela Merkel was, by far, the most frequently named actor in the crisis, followed by the federal government as an entity, and the Federal Minister of Health, Jens Spahn. American President Donald Trump was fourth on the list. Other national actors in the top list included a number of central political figures, but also two interesting deviations: Wolfgang Wodarg (a medical doctor and former politician who publicly criticized the threat by COVID-19 as exaggerated and the political handling as inappropriate and dramatizing a low-risk, flu-like infection) and Christian Drosten (virologist and advisor to the government). The latter two were portrayed by alternative media as antagonists in the debate, signifying a critical heterodox view (Wodarg) versus a mainstream scientific perspective (Drosten) (see also Section \ref{fake_con}). 

Besides Donald Trump, other international politicians on the list included the President of Turkey, Recep Erdoğan, the French President, Emmanuel Macron, and the Austrian Federal Chancellor, Sebastian Kurz. Macron and Kurz were covered as representatives of two relevant, and highly affected, bordering countries; Erdoğan was mentioned frequently in the context of the \emph{Corona and migration} topic (Topic 9, see Section \ref{topics}), consistent with the already mentioned anti-migration sentiments.

The top list also included Swedish climate activist, Greta Thunberg, which was at first surprising because she’s not the most obvious actor to be frequently mentioned in Corona-related news. However, Thunberg has been a frequent object of the coverage in alternative news media before the crisis, as many of them do have a critical stance towards climate change activism, and even a tendency to support climate change conspiracies  \citep{DouglasSutton2015}. It seems plausible that this is a continuation of this negative narrative, which means that links to the pre-existing narrative were constructed to stay consistent with the worldview. For example, some sarcastically mention a “Corona panic” to supplant the “climate panic” (e.g. in Topic 6, Section \ref{topics}). A closer inspection revealed that Thunberg was, for example, criticized for visiting the EU parliament at the beginning of March, despite Corona-related prevention measures already being in place.

Finally, the list of the most frequently mentioned entities also included institutions and group actors. The Robert Koch Institute as the relevant federal agency and research institute for disease control gives recommendations that have binding status in the crisis, so it’s evident that it should show up here. Further, the list included most of the German parties represented in the parliament, but it is interesting to see that the frequency of mentions did not necessarily reflect their relative percentage in the last elections, or their relevance in decisions being made, for that matter. In particular, the right-wing nationalist Alternative für Deutschland (AfD) party scored disproportionately higher than other parties in that respect; it was the sixth most frequently mentioned actor, after the ruling Christian Democrats (CDU), but in front of the second coalition partner, the social democrats (SPD). This does not necessarily reflect notable activities during the crisis, but is rather a manifestation of the right-wing populist viewpoint of many of the alternative news media \citep{frischlichMainstreamAlternativeCoorientation}. So again, this is a continuation of a pre-existing narrative, where the AfD represents a critical viewpoint to the established parties, and the crisis is forged into this general worldview. Like in the case of the Greta Thunberg mentions, a closer inspection supports this interpretation.

\subsubsection {Co-occurence analysis}
\begin{landscape}
\begin{figure}
  \centering
  \includegraphics[width=21.5cm]{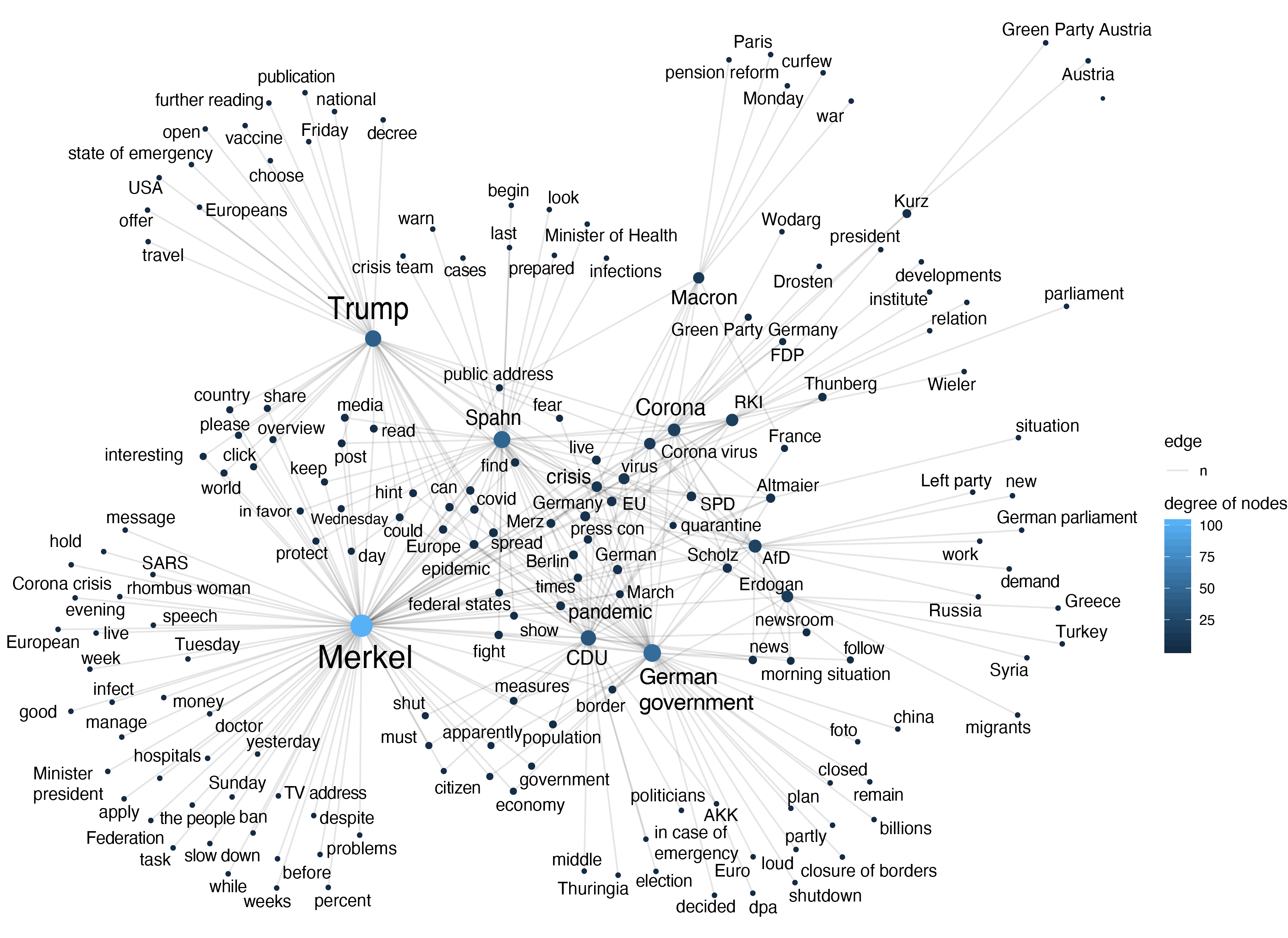}
  \caption{Co-occurrences of most frequent named entities, \emph{N}~=~14,478, filtered by edge count $\geq$ 5}
  \label{fig:coocurrence}
\end{figure}
\end{landscape}

\begin{fontspace}{0.9}{0.9}
The list of named entities, in combination with the previously discussed topics, already offers insight into the reporting and worldview of the alternative news media during the crisis. However, a more detailed analysis of the context in which these actors appear in the messages may help in understanding how they are portrayed and may offer some hints on how the pandemic is framed. For this purpose, we conducted a co-occurrence analysis of the most frequently named entities. In such an analysis, the terms were positioned according to their links to the other (most frequent) terms in the data set, forming a network of more or less densely connected words. This allowed for the identification of strongly connected sub-networks (e.g., if some terms always co-occur with some others in a certain part of the network, but never with terms outside of this substructure) and central terms which co-occur with many other terms (as they are relevant for many topics). In sum, the analysis reveals how meaning is constructed on the level of relating words in given contexts.

An inspection of the resulting co-occurrence graph (see Figure \ref{fig:coocurrence}) unsurprisingly reveals some of the most frequently mentioned entities as being highly connected. Very central and highly connected actors are, for example, Chancellor Angela Merkel, the government, the Minister of Health, Spahn, and the ruling party, CDU. They are connected to multiple terms, which reflects their presence in various contexts, and they are closely linked to a core cluster of words centrally located in the network, which primarily consists of terms describing the domestic situation and the (political) measures against the spread of the virus. In particular, Merkel is seemingly omnipresent and linked to numerous terms from that cluster; in addition, she dominates her own sub-structure of terms referring to her public televised address and her political actions as head of the administration (on the lower left of the network). Some facetious or critical terms either connected to Merkel or the government are also seen, indicating the anti-systemic, heterodox perspective of many alternative news media, for example, “rhombus woman” (a derogatory term referring to Merkel’s signature hand position in public situations).
In contrast to this, other frequently mentioned entities are central to sub-networks that are only weakly linked to the core national structure. International actors like Recep Erdoğan, Sebastian Kurz, Emmanuel Macron, or Donald Trump, are primarily connected to terms describing one specific topic (as also mentioned in Section \ref{topics}), for example, the migration narrative in the case of Erdoğan, or the state of emergency narrative in the case of Trump. These actors \emph{represent} a specific topic. Other actors are connected to a very small number of terms (like Greta Thunberg), and in some cases just one (like the virologist Christian Drosten, who is just linked to the term Corona, as he’s defined by the virus in his function of being a scientific expert).

The co-occurrence analysis supports our previous findings that the alternative news media framed the Corona crisis alongside pre-existing narratives and linked virus-related information to already established concepts (like the migration and refugee debate or climate change activism). Further, their coverage revolves strongly around national topics and systemic criticism, which is in line with the expectations based on the literature (see Section \ref{literaturepopulism}) that identified a strong heterodox, right-wing spin in many alternative news media. Accordingly, the handling of the crisis by political actors is subject to a judgmental analysis and polemic commenting based on viewpoints and patterns established before the pandemic (as visible by using concepts and derogatory terms already introduced during the refugee crisis and before, as in the case of Angela Merkel). So, some of alternative news media's coverage is, indeed, \emph{pandemic populism}.
\end{fontspace}

\clearpage

\begin{landscape}
\begin{table}
 \caption{Overview of conspiracy theories and rumors}
 \centering
  \begin{tabular}{
  >{\raggedright\arraybackslash}p{2.2in}%
  >{\raggedright\arraybackslash}p{1.2in}%
  >{\raggedright\arraybackslash}p{3.8in}%
  >{\raggedleft\arraybackslash}p{0.3in}%
  >{\raggedleft\arraybackslash}p{0.7in}%
  }
    \toprule
    Story & Actors & Narrative & Posts & Interactions \\
    \midrule
Ibuprofen exacerbates progression of COVID-19 infection & Tichys Einblick & Mocking WHO & 1 & 555 \\
 & Sputnik Deutschland \newline ZAROnews & Consideration of different statements & 2 & 83 \\
\midrule
The German federal government already knew about the coming events as early as 2012 and tried to cover it up & MMnews \newline Die Unbestechlichen & Federal government knew about it and could have intervened earlier & 1 & 4 \\
 \midrule
Corona is a man-made laboratory virus & RT Deutsch & Possible accident during research on biological warfare agents & 1 & 2199 \\
\\[-0.8em]
 & Sputnik \newline DWN 
\newline Unzensuriert
\newline Philosophia perennis 
\newline Pravda TV
\newline Krit. Wiss.
\newline Die Unbestechlichen
\newline Politaia & Corona originated in Wuhan close to the national laboratory for biosafety and the laboratory for epidemics and biological weapons & 9 & 1456 \\
\\[-0.8em]
 & Compact & Fact check: Corona mixture of HIV and old corona viruses from Bill Gates’ laboratory? & 1 & 555 \\
 \\[-0.8em]
 & MMnews & Corona as an accident during vaccine development? & 1 & 17 \\
 \\[-0.8em]
 & Pravda TV & Bill Gates funded company brings virus and vaccine into circulation (reference to Rothschild-Epstein-Gates-Darpa connection) & 1 & 17 \\
 \midrule
Corona is not worse than a normal flu epidemic  & NachDenkSeiten \newline Politaia & Drosten vs. Wodarg: Who is right, could there be any truth in Wodarg’s statements? & 3 & 701 \\
\\[-0.8em]
 & DWN \newline Die Unbestechlichen & Corona is a huge fake and is used by politics & 2 & 1039 \\
 & & & & \\
 & Compact \newline MMnews \newline Freie Welt & Citing Wodarg as an opposition: government is scaremongering, Corona no worse than flu & 3 & 940 \\
 \\[-0.8em]
 & Compact & Wodargs statements are correct in parts (mainstream has to agree) & 1 & 508 \\
 \\[-0.8em]
 & Achse des Guten & Collection of arguments against Wodarg & 1 & 304 \\
 & Pravda TV & Corona quite normal wave of flu, vaccinations could aggravate the course of infection & 1 & 36 \\
\bottomrule
    \multicolumn{5}{l}{\emph{Notes:} DWN = Deutsche Wirtschafts Nachrichten, Krit. Wiss. = Kritische Wissenschaft} \\
  \end{tabular}
  \label{tab:Overview_CT}
\end{table}
\end{landscape}

\subsection{Fake news and conspiracy theories}\label{fake_con}

To answer RQ4, we tested the messages included for analysis against data bases of fact-checking institutions to identify totally or partially fabricated news items \citep{quandtFakeNews2019}, which we label as “fake news” here, as well as known conspiracy theories  \citep[following the definition of][]{brothertonDefinitionConspiracyTheory2013}. The analysis did not uncover a large number of posts sharing “fake news” (see Figure \ref{tab:Overview_CT}). We only found three messages covering an explicit “fake news” item, that is, the story of a study proving Ibuprofen to significantly worsen COVID-19. At the time of our analysis, this was a not fully verifiable and dubious rumor that went viral via the WhatsApp messenger service first, and was also covered by many mainstream media sites. While the medical evidence on the association between Ibuprofen and COVID-19 was doubtful, the alleged source of the story (a research institution in Vienna) denied the existence of a study conducted by said institution. Only one of the posts (by \emph{Tichys Einblick}) triggered notable interactions, but overall, the impact and debate regarding the story remained limited. Plausibly, this may be the case as there was a very quick and broad debunking of the story on various news channels, on- and offline, but this remains speculative.

Other news stories can be labeled as conspiracy theories, or rumors fueling uncertainty towards medical and political authorities at best. Our analysis uncovered three major conspiracy theories covered by alternative news media, with a varying factual basis. The first refers to an (actually existing) risk analysis regarding the spread of a hypothetical SARS virus, which had been ordered by the federal government and conducted by the Robert Koch Institute in 2012. Although the risk analysis exists, the framing by the alternative news media gave the story a misleading twist. Conspiracy theorists portrayed the risk analysis as a secret simulation game of the government preempting the actual disease, implying causality and some long-term planning and involvement of the administration. The existence of the paper was not denied by the involved parties, and the conspiracy theory regarding a sinister plan had not caught on---the interactions in our sample remained minimal.

The second conspiracy theory was the most widespread and was covered by multiple alternative news media sites in several variations. The conspiracy theory claims that the virus underlying COVID-19 (SARS-CoV-2) is man-made and was purposefully designed in a laboratory. In most cases, the suspected production facility is a national lab for bio security or weapons in Wuhan. Some also portray the origins of the virus as an accident while developing bio weapons or a vaccine. One medium also speculated about a company financed by Bill Gates to be behind the virus, linking its outbreak to a wider conspiracy network. These speculations about the origin of the virus were not only widely covered by alternative news media, but also triggered a high number of interactions, with one specific news item by the German branch of \emph{Russia Today} (RT) receiving more than 2000 interactions.

The final group of messages analyzed was a very specific one based on the public statements of a politician and physician, Wolfgang Wodarg, who is portrayed in the general public debate and political action as being devoid of a factual basis as he claimed that the risks for negative outcomes due to COVID-19 for the general population would be marginal. He framed this notion in a way that implied panic mongering as a main driver of the crisis. Conspiracy theorists used Wodarg’s statements as a reference to claim that the pandemic would be a major “fake.” Some of the alternative media turned this into a “duel” between Wodarg as one legitimate opinion, and the other one being one of the country’s leading virologist, Christian Drosten, who served as an advisor of the government during the pandemic and publicly countered Wodarg’s statements. The personalizing of the conflict served alternative news media’s general narrative; Wodarg symbolizes their heterodox viewpoint opposing “the system,” while Drosten symbolizes the orthodox perspective of the administration and the societal mainstream. Notably, some mainstream media relied on the same antagonistic perspective and picked up the topic in a similarly personalized way. The group of Wodarg-related messages elicited a comparably high number of interactions, potentially indicating a larger interest by the audience and an accordingly higher potential impact (naturally, this would need to be confirmed by survey studies).

Overall, the posts that could be linked to disinformation and conspiracy theories only made up 1.1\% of the overall data set and 1.1\% of the interactions. Although our analyses reflect governmental reports about circulating narratives, such as those by the European external action service \citep{eeasShortAssessmentNarratives202004}, our data provides nuance to the debate by showing that the overall number is, in contrast to public fears, relatively low. However, when analyzing the data set, we found many of these posts and messages show up in a secondary distribution system, for example, the story on the risk analysis of the German government posted by MMnews on January 29 was taken up by several conspiracy theorists’ YouTube channels (e.g., Schrang TV) and AfD members (e.g., Martin Sichert and Alice Weidel) \citep{CTyoutube} and thereby the document created over 15,000 interactions and potentially reached over 2.5 million users solely via Facebook. These channels, in turn, took the posts as reference for their own worldviews; through relaying the messages into a secondary system, alternative news' messages became credible sources and proof for a deviant, heterodox perspective. For instance, a large-scale content analysis of a popular right-wing YouTube channel showed that alternative news media were used frequently as evidence for the credibility of the worldview transmitted there \citep{frischlichMainstreamAlternativeCoorientation}. In that sense, their potential influence and contribution to societal confusion may be indirect.

\section{Conclusion}\label{conclusion}
As noted in the introductory section, alternative news media were suspected of being problematic, confusing, and misleading voices in the Corona crisis by politicians, journalists, and the public alike. Some politicians, even in liberal democracies, called for legal action to prevent the public from being dangerously misinformed by channels that propagate rumors or lies, which in turn, were feared to fuel the (self)harm of citizens, and indeed, some countries imposed stricter rules and regulations based on these assumptions.

The current analysis of the Facebook activities of German language alternative news media, covering the beginning of January until mid-March, paints a more nuanced picture of the situation. Our analysis does not reveal a notable amount of “fake news” (as identified by fact checkers) in these channels. In that sense, alternative news media were not a major force spreading outright lies or fabrications during that time, and they were not openly conspicuous. Given other research and theorizing on fake news and disinformation \citep{quandtFakeNews2019}, this is not surprising; the best propaganda is disguised \citep{jowettPropagandaPersuasion2012} and the best lie always has a true core.

Alternative news media outlets like those studied here have been found to embed their ideologically-biased content subtly into an overall seemingly innocuous communication strategy \citep{frischlichMainstreamAlternativeCoorientation}. Broad and open disinformation strategies would not fit the way these alternative news media sites operate. These outlets are, instead, characterized by commenting and criticizing the orthodox, majority perspective, while at the same time depending on the more traditional mainstream media as a necessary counterpart \citep{figenschouChallengingJournalisticAuthority2019}. In some ways, alternative news media function much like a photo negative of what and how the large mainstream media reports on; the outlines are the same, but they are mirrored with reversed colors.

Our findings show that the alternative news media remain true to their principles during the examined stage of the Corona crisis. The large majority of posts mirrored traditional mainstream media reports in terms of their topical structure and the actors involved. We did not see a completely fabricated reality here. However, as expected, we did see a certain populist spin in their posts \citep{devreesePopulismExpressionPolitical2018}, that is, an anti-establishment tone, critical of public institutions and political actions of the administration, evoking strong emotions. So, while the topics did mirror the mainstream, the tone of the posts did not. The coverage was biased toward focusing on criticism regarding the communication and management of politicians and mainstream media, often by changing the framing of the news to fit their generally “anti-establishment” perspective. As such, the alternative news media examined here used COVID-19-related information to foster their long-term narratives, namely a critical stance toward established politicians \citep{frischlichMainstreamAlternativeCoorientation}, refugees and immigration \citep{heftBreitbartComparingRight2019} or, more recently, a tendency towards climate change conspiracies \citep{DouglasSutton2015}. In that sense, the Corona crisis does not change the general logic of German alternative news media’s coverage, but it shows how world events are adjusted and assimilated to their respective ideology.

Nevertheless, our data also shows that alternative news media did prominently feature rumors and conspiracy theories regarding COVID-19 and the origins of SARS-CoV-2 and created a notable amount of interactions with this kind of content. Again, these stories were strongly linked to alternative news media's general worldview and pre-existing narratives, like a strong opposition to climate activism (as visible in the claim that the Corona crisis is just following the patterns of panic mongering introduced by climate activists), anti-vaccination attitudes (manifesting in the allegation that Bill Gates brought the virus into circulation to make money on the vaccine), or disapproval of foreigners (with the respective conspiracy theory that the Chinese developed the virus as a weapon and spread it via tourists and immigrants to destroy the West). In essence, our analysis revealed alternative news media to indeed report on the crisis in the form of \emph{pandemic populism}, following a specific and expected communication style \citep{devreesePopulismExpressionPolitical2018}, but did not simply spread open lies.

It is noteworthy that this information mix is much more likely to contribute to the feared “infodemic” \citep{who_novel_2020} as media users are usually better in noting highly non-credible sources and disinformation bits \citep{pennycookPriorExposureIncreases2018, pennycookFightingMisinformationSocial2019}. The observed recontextualization of information, plausibly with the goal of undermining trust in public statements and policies \citep{eeasShortAssessmentNarratives202004}, and the adaption of events in alternative news media’s general, anti-systemic meta-narrative, is likely to sow uncertainty and confusion. And uncertainty can, in turn, benefit authoritarian worldviews  \citep{duckittImpactSocialThreat2003} and even extremist positions \citep{riegerPropagandaInsecureUnstructured2017}.

Further, the larger impact of alternative news media is likely to be indirect. As the data analysis shows, direct interactions on the basis of alternative news media is high, and we found several cases where their coverage was shared and picked up elsewhere, for example, on conspiracy theorists’ Youtube channels, which serve as a secondary distribution and amplification system. Here, the messages of alternative news media are referenced as sources and critical voices that speak the truth (in contrast to professionalized mainstream media).

Our analysis has notable limitations, though, that need to be considered. First, we focused on a single, specific country (Germany), a relatively short time frame (initial phase of spreading and first pandemic phase), and examined only the Facebook activities of the most relevant alternative news media within the country. A subsequent analysis will shed some light on the parallel Facebook activities of mainstream media (in an upcoming working paper of the authors). Naturally, comparative data from other countries, further social media channels, and a longer time period, might be helpful in verifying what we deduced as more general principles from the given case, and in particular, the information strategies of alternative news media in the Corona crisis.

As it is, this initial analysis of Facebook posts covering the crisis does not support some of the public fears about a widespread dissemination of completely fabricated news through the alternative news media. Nevertheless, alternative news media contribute to public confusion by their specific framing strategies and by constructing a contradictory, menacing, and distrusting worldview, which calls any official statement into question. This may be dysfunctional for social coherence in an unprecedented time of crisis; still, such communication largely operates within the given limits of discourse in a liberal democracy. Moreover, studies on the association between the consumption of alternative news media and risky behavior during the COVID-19 pandemic are missing so far. Thus, while we know more about the content through the present study, the actual effects on the users are just (plausible) conjectures.

Some of the more radical political demands of limiting or censoring such debates seem to be unwarranted on the basis of our preliminary analysis; it is even likely that such legal action may indeed be counterproductive because it would further feed anti-systemic sentiments, potentially without having the intended short-term effects. It may also be argued that especially in times of crisis, the strength of a democracy is first and foremost reflected in its open and plural debates, even if some of it is not particularly helpful or even hostile, as documented in our preliminary analysis of alternative news media's coverage during the Corona crisis.

\clearpage

\bibliography{references}

\end{document}